\documentclass[final,5p,times,twocolumn]{elsarticle}

\usepackage{amsmath,amsfonts,amssymb,graphicx,color}

\usepackage[labelfont=bf,font=small]{caption}

\biboptions{sort&compress}

\journal{Physics Letters B}

\begin{document}

\begin{frontmatter}

\title{Charmonium Spectral Functions and Transport Properties of 
Quark-Gluon Plasma}

\author[ITPGU]{Si-xue Qin}
\ead{sixueqin@th.physik.uni-frankfurt.de}

\author[ITPGU]{Dirk H.\ Rischke}
\ead{drischke@th.physik.uni-frankfurt.de}

\address[ITPGU]{Institute for Theoretical Physics, Goethe University, 
Max-von-Laue-Str.\ 1, D-60438 Frankfurt am Main, Germany}

\date{\today}

\begin{abstract}
We study vacuum masses of charmonia and the charm-quark diffusion 
coefficient in the quark-gluon plasma based on the spectral
representation for meson correlators. To calculate the correlators, 
we solve the quark gap equation and the
inhomogeneous Bethe-Salpeter equation in the rainbow-ladder
approximation. It is found that the ground-state masses of charmonia 
in the pseudoscalar, scalar, and vector channels can be well described. 
For $1.5\,T_c<T<3.0\,T_c$, the value of the diffusion coefficient $D$ 
is comparable with that obtained by lattice QCD and experiments: 
$3.4<2\pi TD<5.9$. Relating the diffusion coefficient with 
the ratio of shear viscosity to entropy density $\eta/s$ of the
quark-gluon plasma, we obtain values in the range $0.09<\eta/s<0.16$.
\end{abstract}


\end{frontmatter}

The charmonium system is a bound state of a charmed quark--anti-quark
pair. The first charmonium state $J/\psi$ was found simultaneously at 
BNL \cite{Aubert:1974js} and at SLAC \cite{Augustin:1974xw} in 1974. 
Charmonium spectroscopy plays the same role 
\cite{Voloshin:2007dx,Wiedner:2010zz} for
understanding the strong interaction, described by quantum
chromodynamics (QCD), as does the spectroscopy of
positronium or of the hydrogen atom 
for the electromagnetic interaction, described by quantum
electrodynamics (QED). 

Charm quarks are also produced in hard parton interactions in the early stage of 
heavy-ion collisions, e.g.\ at the Relativistic Heavy Ion Collider (RHIC)
and the Large Hadron Collider (LHC). During the further evolution
of the fireball, these quarks interact with the quark-gluon plasma
(QGP) created in such collisions. The ensuing loss of energy of a charm quark is
different from the one experienced by light quarks 
\cite{Wang:1991xy,Braaten:1991we}. A comparison
between the energy loss for light quarks with that for heavy ones can
provide insight into properties of the QGP. 

Even in a hot and dense medium, charm quarks can form bound states
with other light or heavy quarks. The formation and dissociation
of these states depends on the properties of the surrounding
medium. For instance, it was proposed \cite{Matsui:1986dk}
that, due to color screening, the formation of $J/\psi$ is suppressed 
in the QGP, which can serve as a signal of the deconfinement phase transition. 
More recent calculations within lattice QCD
\cite{Datta:2003ww,Asakawa:2003re,Aarts:2007pk,Ohno:2011zc}, 
however, show that the $J/\psi$ may actually survive up to temperatures
exceeding the critical temperature $T_c$ of the deconfinement and
chiral phase transition.
Therefore, it is an interesting and meaningful task to understand 
charmonium properties in vacuum and medium systematically. 

A first-principle method to study charmonium properties is lattice
QCD. Within this approach, the charmonium spectrum, including ground,
excited, and exotic states, has been computed at zero temperature, $T=0$
\cite{Davies:1995db,Okamoto:2001jb,Dudek:2007wv}, finding rather
good agreement with experimental data. Transport properties,
e.g., the charm quark diffusion coefficient, which are closely related to
charmonium spectral functions, are also calculable within lattice QCD
\cite{Banerjee:2011ra,Ding:2012sp}. 
The charm quark diffusion coefficient has also been studied within 
a $T$-matrix approach \cite{Cabrera:2006wh} and a 
relativistically covariant approach based on QCD sum rules 
\cite{Gubler:2011ua}.

Assuming that the interaction between charm quarks can be described 
by a potential, one can adopt nonrelativistic potential models 
to study charmonium properties \cite{Eichten:1979ms}. 
The parameters of the potential can be adjusted to the vacuum
charmonium spectrum. In order to study charmonia at
nonzero temperatures, one can generalize the vacuum
potential to a temperature-dependent one based on models
\cite{Mocsy:2008eg} or lattice-QCD results \cite{Kaczmarek:2005ui}. 

Dyson-Schwinger equations (DSEs) \cite{Roberts:1994dr,Roberts:2007ji} 
which include both dynamical chiral symmetry breaking and confinement 
serve as a nonperturbative continuum approach for studying QCD. 
At $T=0$, DSEs have been used to study properties of bound
states, e.g., the light and heavy hadron spectrum 
\cite{Chang:2011ei,Blank:2011ha}, pion properties 
\cite{Chang:2013nia,Chang:2013pq}, as well as hadron form factors 
\cite{Maris:2003vk,Chang:2011vu}. At $T\ne 0$, the chiral and
deconfinement transitions and the excitations in the QGP have been studied 
through solving the quark gap equation \cite{Qin:2010pc,Qin:2013ufa}. 
Recently, a novel spectral representation has been developed to study 
in-medium hadron properties and the electrical conductivity of the QGP
\cite{Qin:2013aaa}. These results are consistent with experiment and
lattice QCD, which establishes the DSE approach as a powerful and
reliable tool to study the properties of hadrons and 
strong-interaction matter. 

In this work, we employ DSEs to study charmonium spectral functions
and transport properties of the QGP.  
First, we calculate the charmonium spectrum at $T=0$ in order to 
fix the charm quark current mass. Then, we study the charm quark 
number susceptibility (QNS) and diffusion coefficient at $T>T_c$. 
At last, we use a formula obtained by perturbation theory to translate
the diffusion coefficient to the ratio of shear viscosity to entropy 
density, i.e., $\eta/s$, of the QGP.

In the imaginary-time formalism of thermal field theory \cite{LeBellac:2000wh}, 
the Matsubara correlation function of a local meson operator
$J_H(\tau,\vec{x}\,)$ is defined as
\begin{eqnarray}
\Pi_H(\tau,\vec{x}\,)=\langle J_{H}(\tau,\vec{x}\,)
J_{H}^\dagger(0,\vec{0}\,) \rangle_{\beta}\;,
\label{eq:msncor}
\end{eqnarray}
where $\beta=1/T$, $\tau$ is the imaginary time with $0<\tau<\beta$, 
and $\langle \ldots \rangle_\beta$ denotes the thermal average. 
The operator $J_H$ has the following form
\begin{eqnarray}
J_H(\tau,\vec{x}\,)=\bar{q}(\tau,\vec{x}\,)\gamma_H q(\tau,\vec{x}\,)\;,
\end{eqnarray}
with $\gamma_H=\mathbf{1},\gamma_5,\gamma_\mu,\gamma_5\gamma_\mu$ 
for scalar, pseudo-scalar, vector, and axial-vector channels, respectively. 

In terms of Green functions the meson correlation functions are defined as
\begin{eqnarray}
\parbox{72mm}{\includegraphics[width=\linewidth]{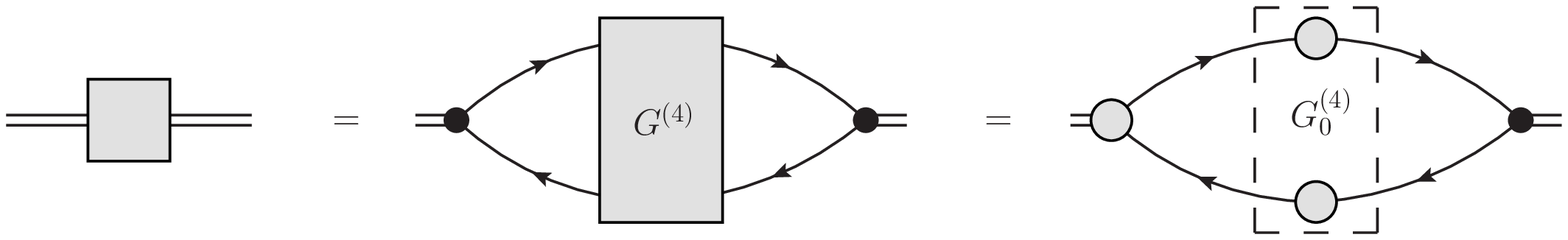}}\;,
\label{eq:cor}
\end{eqnarray}
where gray circular blobs denote dressed propagators $S$ and 
vertices $\Gamma_H$, $G^{(4)}$ denotes the full quark--anti-quark 
four-point Green function, $G_0^{(4)}$ denotes the two 
disconnected dressed quark propagators in the dashed box, and
black dots denote bare propagators or vertices. 

The dressed quark propagator $S$ is a solution of the 
quark gap equation which reads
\begin{eqnarray}
\parbox{65mm}{\includegraphics[width=\linewidth]{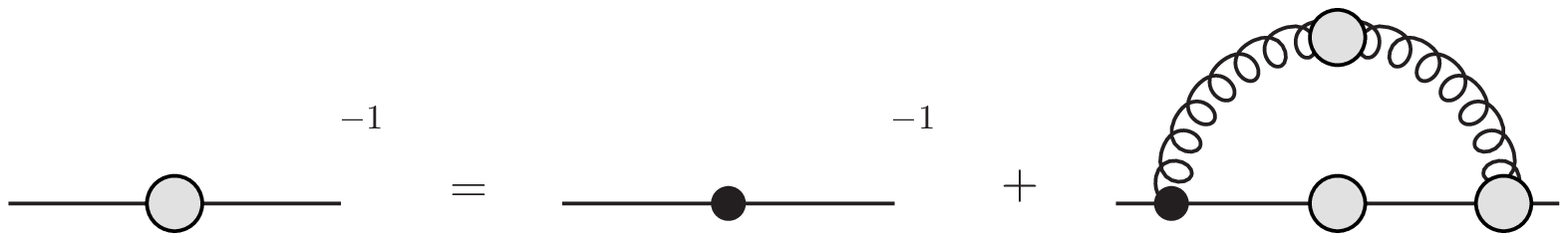}}\;.
\label{eq:gap}
\end{eqnarray}
The dressed quark propagator depends on the dressed gluon 
propagator $D_{\mu\nu}^{ab}$ and the dressed quark-gluon vertex 
$\Gamma_\mu^a\,$. The dressed quark propagator $S(\tilde\omega_n,\vec{p}\,)$ can be generally decomposed as
\begin{eqnarray}
S=1/[i\vec{\gamma}\cdot\vec{p}\,
A(\tilde\omega_n^2,\vec{p}\,^2) + i\gamma_4\tilde\omega_n\,
C(\tilde\omega_n^2,\vec{p}\,^2)+ B(\tilde\omega_n^2,\vec{p}\,^2)]\;,\notag
\end{eqnarray}
where $\tilde\omega_n=(2n+1)\pi T,\, n\in Z$, are the fermionic
Matsubara frequencies, and $A,B$, and $C$ are scalar functions. 
The mass scale of quarks can be defined as 
\begin{eqnarray}
M_0=\frac{B(\tilde\omega_0^2,\vec{0}\,)}{C(\tilde\omega_0^2,\vec{0}\,)} \;.
\end{eqnarray} 

The dressed vertex $\Gamma_H$ satisfies the inhomogeneous 
Bethe-Salpeter equation (BSE),
\begin{eqnarray}
\parbox{62mm}{\includegraphics[width=\linewidth]{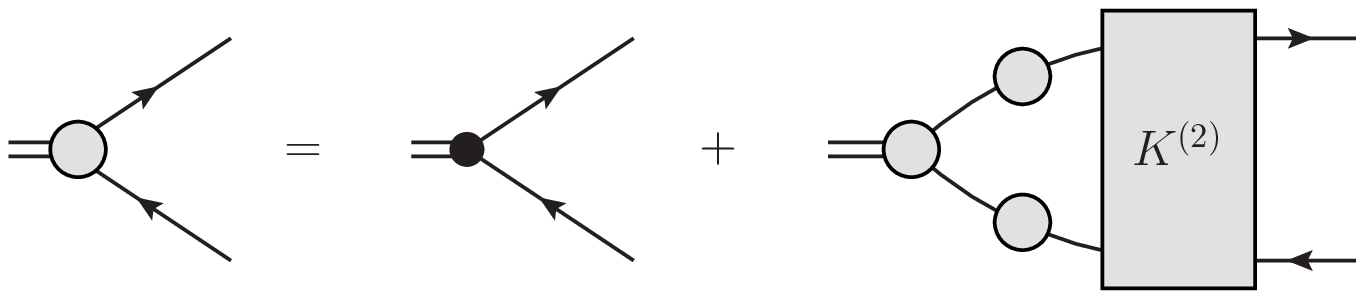}}\;,
\label{eq:bse}
\end{eqnarray}
where $K^{(2)}$ denotes the two-particle irreducible kernel. 
The above equation needs the dressed quark propagator as input. 
Its solution can be decomposed according to the 
$J^{P}$ quantum number of the corresponding meson channel $H$.

Inserting the solutions of Eqs.\ \eqref{eq:gap} and \eqref{eq:bse}
into Eq.\ \eqref{eq:cor}, we obtain the imaginary-time 
charmonium correlation functions. However, for solving Eqs.\
\eqref{eq:gap} and \eqref{eq:bse}, we have to specify
$D_{\mu\nu}^{ab}$, $\Gamma_\mu^a$, and $K^{(2)}$. To this end, 
we use the abelian rainbow-ladder (RL) approximation. 
The rainbow part of this approximation
consists of (color indices are suppressed)
\begin{equation}
g^2 D_{\mu\nu}(k_\Omega)\Gamma_\nu(\tilde\omega_n,\vec{p}\,;\tilde\omega_l,\vec{q}\,)
=D^{\rm eff}_{\mu\nu}(k_\Omega)\gamma_\nu\;,
\end{equation}
with the effective gluon propagator written as
\begin{equation}
D^{\rm eff}_{\mu\nu}(k_\Omega)=P^T_{\mu\nu}\mathcal{D}(k_\Omega^2)
+P^L_{\mu\nu}\mathcal{D}(k_\Omega^2+m_g^2)\;,
\end{equation}
where $k_\Omega=(\tilde\omega_n-\tilde\omega_l,\vec{p}\,-\vec{q}\,)$,
$P^{T,L}_{\mu\nu}$ are transverse and longitudinal projection tensors,
respectively, $\mathcal{D}$ is the gluon dress function 
which describes the effective interaction, and $m_g$ is the 
gluon Debye mass. The ladder part of the RL approximation
is given by
\begin{eqnarray}
K^{(2)}(\tilde\omega_n,\vec{p}\,;\tilde\omega_l,\vec{q}\,)=
D^{\rm eff}_{\mu\nu}(k_\Omega)[(i\gamma_\mu) \otimes (i\gamma_\nu)]\;,
\end{eqnarray}
which expresses the two-particle irreducible kernel in terms of one-gluon 
exchange. Note that the RL approximation is the leading term 
in a symmetry-preserving approximation scheme. The solutions of 
Eqs.\ \eqref{eq:gap} and \eqref{eq:bse} satisfy Ward-Takahashi 
identities \cite{Ball:1980ay,Qin:2013mta,Qin:2014vya}. 

Now the quark gap equation and the inhomogeneous BSE, i.e., 
Eqs.\ \eqref{eq:gap} and \eqref{eq:bse}, can be self-consistently 
solved once the gluon dress function $\mathcal{D}$ is given. 
Here, the modern one-loop renormalization-group-improved 
interaction model \cite{Qin:2011dd,Qin:2011xq} is adopted. This
model has two parameters: a width $\xi$ and a strength $d$. 
With the product $\xi d$ fixed and $\xi\in[0.4,0.6]$ GeV, one can 
obtain  a uniformly good description of pseudoscalar and vector mesons
in vacuum with masses $\lesssim1$ GeV. We use $\xi=0.5$ GeV. 
At $T>T_c$, we introduce a Debye mass $m_g$ in the longitudinal part 
of the gluon propagator and a logarithmic screening for the 
nonperturbative interaction \cite{Qin:2013ufa,Qin:2013aaa} 
in order that physical quantities, e.g., the thermal quark masses 
for massless quarks and the electrical conductivity of the QGP, 
are consistent with lattice QCD \cite{Karsch:2009tp,Amato:2013naa}.

All information which we are interested in is embedded in the 
spectral function of the charmonium correlation function. 
In energy-momentum space, the spectral function is defined 
as the imaginary part of the retarded correlation function,
\begin{eqnarray}
\rho_H(\omega,\vec{p}\,)&=&2\,{\rm Im}\,\Pi_H^R(\omega,\vec{p}\,)\notag\;,\\
&=&2\,{\rm Im}\Pi_H(i\omega_n,\vec{p}\,)|_{ i\omega_n\to\omega+i\epsilon}\;,
\label{eq:img}
\end{eqnarray}
where $\omega_n=2n\pi T,\, n\in Z$, are the bosonic Matsubara frequencies. 
Then, the spectral representation at zero momentum ($\vec{p}\,=\vec{0}\,$) reads
\begin{eqnarray}
\Pi_H(\omega_n^{2})=\int_{0}^{\infty}\frac{d\omega^2}{2\pi} 
\frac{\rho_H(\omega)}{\omega^2 +\omega_n^2}-({\rm
subtraction})\;,
\label{eq:spec1}
\end{eqnarray}
where the dependence of $\Pi_H$ on $\omega_n$ is quadratic, since it
is an even function of $\omega_n$ at $\vec{p}\,=\vec{0}\,$.
An appropriate subtraction is required because the spectral integral
does not converge for meson correlation functions, i.e., 
$\rho_H(\omega\to\infty)\propto \omega^2$. This divergence manifests itself
in a corresponding divergence of the loop integral in Eq.\ \eqref{eq:cor}. 

In order to solve the divergence problem, one of us (S.Q.) has
suggested to introduce a discrete transform for $\Pi_H$ \cite{Qin:2013aaa},
\begin{eqnarray}
\hat\Pi_H(\omega_{n_1}^{2},\omega_{n_2}^2,\omega_{n_3}^2)=
\sum_{i=1}^{3}\Pi_H(\omega_{n_i}^2) 
\prod_{j\ne i}^{3}\frac{1}{\omega_{n_i}^2-\omega_{n_j}^2}\;,
\label{eq:transform}
\end{eqnarray}
where $\omega_{n_1}\ne\omega_{n_2}\ne\omega_{n_3}$. Then, we can obtain 
the well-defined spectral representation
\begin{equation}
\begin{split}
\hat\Pi_H(\omega_{n_1}^{2},\omega_{n_2}^2,\omega_{n_3}^2)=
\int_{0}^{\infty}\frac{d\omega^2}{2\pi} \, \rho_{H}(\omega) \prod_{i=1}^{3}
\frac{1}{\omega^2+\omega_{n_i}^2}\;.
\label{eq:spec}
\end{split}
\end{equation}
By introducing a one-variable correlator as 
$\tilde\Pi_H(\omega_{n}^{2})=\hat\Pi_H(\omega_{n}^2,\omega_{n+1}^2,
\omega_{n+2}^2)$, we can reduce Eq.\ \eqref{eq:spec} to a 
one-dimensional equation for numerical convenience.

At $T=0$, the Matsubara frequencies become continuous
and the system has an $O(4)$ symmetry. Then, the spectral
representation \eqref{eq:spec} can be written in an $O(4)$ covariant
form. The charmonium spectrum in vacuum can be extracted from the 
spectral function $\rho_H(\omega)$. 

At $T\ne0$, especially, $T>T_c$, a charmonium state, 
e.g., $J/\psi$, can survive up to a certain temperature above which it
dissolves. With increasing $T$, the corresponding peak in the spectral
function becomes broader and decreases in height, and finally
disappears altogether.
Qualitatively, we may identify the charmonium dissociation temperature
as the temperature where the charmonium peak in the spectral function 
becomes indistinguishable from the background.

Of great interest are transport properties of charm quarks 
because they reflect the dynamics of the QGP. Several transport
coefficients can be extracted from the electromagnetic (i.e., vector)
current correlation function $\Pi_V^{\mu\nu}$. First, the QNS is defined as
\begin{eqnarray}
\chi_{00}=\frac{\partial n(\mu,T)}{\partial\mu}\bigg|_{\mu=0}=\Pi_V^{44}(0)\;,
\label{eq:chi}
\end{eqnarray}
where $n(\mu,T)$ is the quark number density. The second equality is 
derived from the vector Ward identity -- a result of vector-current
conservation. According to Eq.\ \eqref{eq:cor}, $\Pi_V^{44}$ only 
depends on the dressed vector vertex and quark propagator, i.e., 
$\Gamma_V^\mu$ and $S$. In the non-interacting 
quasi-particle approximation, $\Gamma_V^\mu=\gamma^\mu$ and 
$S=1/[i\vec{\gamma}\cdot\vec{p}\, + i\gamma_4\tilde\omega_n + M_0]$. 
Then, $\chi_{00}$ is given by
\begin{eqnarray}
\chi_{00}=-4N_{c}\int\frac{d^3\vec{k}}{(2\pi)^{3}}
\frac{\partial n_f(E_{\vec{k}})}{\partial E_{\vec{k}}}\;,
\label{eq:chiq}
\end{eqnarray}
where $E_{\vec{k}}=\sqrt{M_0^2+\vec{k}\,^2}$ is the 
quasi-particle energy, and $n_f$ is the Fermi distribution function.

Without the quasi-particle approximation, we have to
evaluate the QNS numerically. To this end,
we turn to the spectral representation Eq.~\eqref{eq:spec}. From Eqs.\
\eqref{eq:spec1} and \eqref{eq:chi}, 
the zeroth component of the vector spectral function can be written as
\begin{eqnarray}
\rho_V^{00}(\omega)=2\pi\chi_{00} \omega\delta(\omega)\;.
\end{eqnarray}
Inserting this into Eq.\ \eqref{eq:spec}, we have
\begin{eqnarray}
\chi_{00}=\hat\Pi_V^{44}(\omega_{n_1}^2=0,\omega_{n_2}^2,\omega_{n_3}^2)\, \omega_{n_2}^2 \,\omega_{n_3}^2\;,\
\label{eq:chifull}
\end{eqnarray}
which is well-defined and equivalent for any $\omega_{n_2}\ne \omega_{n_3}\ne
0$. In comparison to Eq.\ \eqref{eq:chi}, the above expression 
is much easier to calculate numerically.

\begin{figure}
\centering
\includegraphics[width=0.95\linewidth]{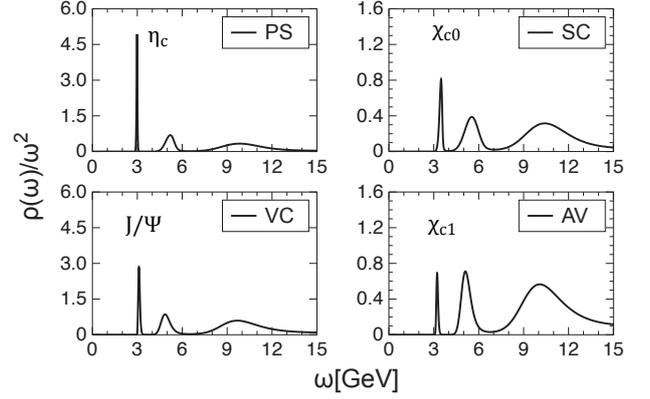}
\caption{Charmonium spectral functions at zero temperature.} \label{fig:zerospec}
\end{figure}

\begin{table}
\centering
\caption{The Euclidean constituent mass of the charm quark and charmonium masses at zero temperature 
($m^\zeta_c=0.79$GeV, $\zeta=19$GeV, and dimensional quantities are given in GeV).
\label{tab:charmspec}}
\begin{tabular*}{1.0\linewidth}{@{\extracolsep{\fill}}cccccc}\hline
~ & $m^E_c$ & $\eta_c$ & $J/\psi$ & $\chi_{c0}$ & $\chi_{c1}$ \\\hline
this work & 1.396 & 2.980 & 3.113 & 3.476 & 3.227 \\
PDG & 1.275 & 2.980 & 3.097 & 3.415 & 3.510 \\\hline
\end{tabular*}
\end{table}

From the Kubo formula, the heavy quark diffusion coefficient can be expressed as
\begin{eqnarray}
D=\frac{1}{6\chi_{00}}\lim_{\omega\to 0}
\sum_{i=1}^{3}\frac{\rho_V^{ii}(\omega)}{\omega}\;,
\label{eq:hqd}
\end{eqnarray}
where $\rho_V^{ii}$ are the spatial components of the vector spectral 
function (in what follows, the summation is suppressed unless
stated). In Ref.\ \cite{Moore:2004tg}, Moore and Teaney perturbatively
calculated the ratio of $D$ to the transport coefficient 
$\eta/(\epsilon+p)$, where $\epsilon$ is the energy density and $p$ is
the pressure. It is found that for a QGP with two light flavors, the ratio 
is around $6$ and has a weak dependence on the coupling strength. 
Using the thermodynamic identity $\epsilon+p=sT$ (at zero chemical
potential), we can translate $D$ to $\eta/s$ as
\begin{eqnarray}
\frac{\eta}{s} \approx \frac{1}{6}TD\;.
\label{eq:etas}
\end{eqnarray}

In order to extract the spectral function from $\hat\Pi_H$ 
which is given by the solution of the DSE, we adopt the maximum
entropy method (MEM)
\cite{Bryan:1990tv,Nickel:2006mm,Mueller:2010ah}. At $T=0$, we follow 
lattice-QCD studies \cite{Asakawa:2000tr} and choose the MEM default
model as $m_{\rm fr} \omega^2$, where the coefficient $m_{\rm fr}$ is
calculated in the non-interacting limit
\cite{Karsch:2003wy,Aarts:2005hg}. The results are shown in Fig.\ 
\ref{fig:zerospec}. In each channel, the first peak is sharp. 
This means that the ground-state signal is strong. The corresponding masses 
are listed in Table \ref{tab:charmspec}, where the pseudoscalar 
channel $\eta_c$ is fitted by adjusting the current charm quark mass
$m_c^\zeta$, and the Euclidean constituent mass of the charm quark $m^E_c:=\{\sqrt{p^2}\, |\, p^2>0, {p^2}=B^2(p^2)/A^2(p^2)\}$ (because of the $O(4)$ symmetry at $T=0$, $A=C$ are functions of four-momentum squared $p^2$). It is found that, with the exception of the axial-vector
channel, all masses agree well with their experimental values. 
The $\chi_{c1}$ mass (and similarly the $a_1$ mass) comes out noticeably
smaller than the experimental value, 
because the RL approximation misses the anomalous 
chromomagnetic effect in the quark-gluon vertex \cite{Chang:2010hb}. 
Going beyond the RL approximation \cite{Chang:2011ei,Chang:2011vu}, 
one can remedy this drawback. Nevertheless, it is safe to use the 
RL approximation for the vector channel.

\begin{figure}
\centering
\includegraphics[width=0.95\linewidth]{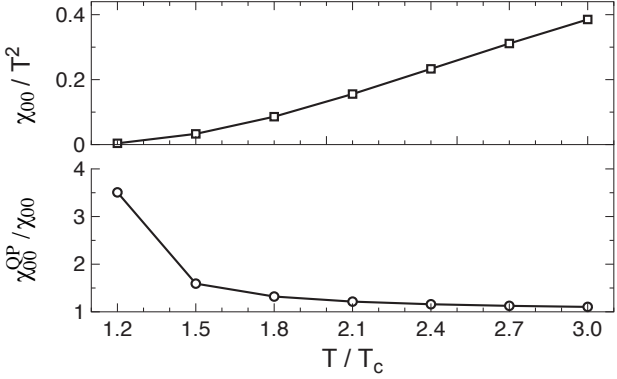}
\caption{The heavy quark number susceptibility 
(the upper panel) and the ratio to that obtained in 
the quasi-particle picture (the lower panel) as a function of
temperature.} 
\label{fig:chi00}
\end{figure}

Next, we study the QNS of charmed quarks in the QGP. 
The result calculated by Eq.\ \eqref{eq:chifull} (denoted by
$\chi_{00}$) is shown in the upper panel of Fig.\ \ref{fig:chi00}. 
For comparison, we also present the result obtained by the
quasi-particle formula \eqref{eq:chiq} (denoted by $\chi_{00}^{\rm
  QP}$). As illustrated in the lower panel of Fig.\ \ref{fig:chi00}, 
the ratio $\chi_{00}^{\rm QP}/\chi_{00}$ decreases with increasing 
temperature, which means that the quasi-particle picture works well
at high temperature, but fails in the neighborhood of $T_c$.

\begin{figure}
\centering
\includegraphics[width=0.95\linewidth]{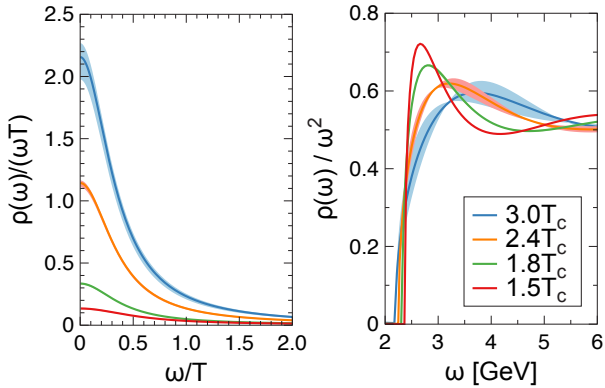}
\caption{The vector spectral function at different temperatures, 
where the shaded regions around the curves (produced by changing the
MEM default model) provide an estimate for the systematic error.} \label{fig:rhoall}
\end{figure}
At $T>T_c$, it is assumed that the vector spectral function has 
two parts, i.e., a low-energy transport peak and a 
continuous part above $2M_0$. Then, we prepare the default model as
\begin{eqnarray}
\rho_V^{ii}(\omega) &= & \frac{6\chi_{00}T}{M_0}\frac{\omega\eta_D}{
\omega^2+\eta_D^2} + \frac{3}{2\pi}\Theta\left(\omega^2-4M_0^2\right)\notag\\
&& \times~\omega^2\tanh\left({\omega}/{4T}\right)\sqrt{1-{4M_0^2}/{\omega^2}}\notag\\
&& \times~\left[1+{4M_0^2}/{\omega^2}\right]\;,
\end{eqnarray}
where $\eta_D=\frac{T}{M_0D}$ is the drag coefficient. Note that, 
as for $T=0$, there is no resonance peak specified in the continuous part. 

\begin{figure}
\centering
\includegraphics[width=0.95\linewidth]{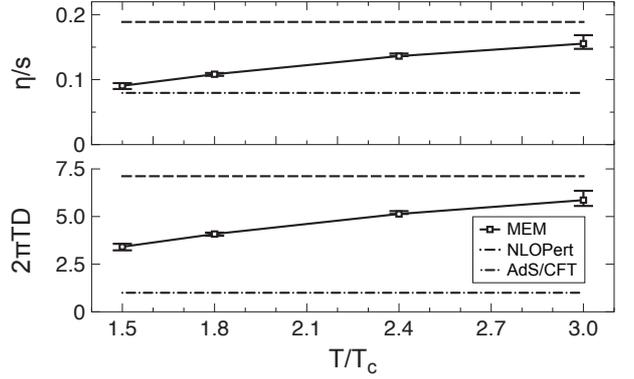}
\caption{The ratio $\eta/s$ and the heavy quark diffusion coefficient 
as a function of temperature. The dashed line is obtained by an NLO
perturbative calculation at $T=3T_c$, 
and the dot-dashed line is given by AdS/CFT. 
The error bars are obtained by altering the default model.} \label{fig:trans}
\end{figure}
In order to completely determine the default model, we need prior
information on $\eta_D$. In the neighborhood of $T_c$, i.e., $T\gtrsim
T_c$, we assume the system to be a strongly coupled QGP, where the
AdS/CFT correspondence gives $\eta_D=\frac{2\pi T^2}{M_0}$ \cite{Kovtun:2003wp}. 
At high temperature, e.g., $T\gtrsim 3T_c$, perturbative QCD at 
next-to-leading order (NLO) yields the thermal quark mass (for 
massless quarks) $m_T=\frac{g_s
  T}{\sqrt{6}}\left(1+1.867\frac{g_s}{4\pi}\right)$ 
\cite{Carrington:2008dw} and the drag coefficient 
$\eta_D=\frac{8\pi T^2}{3M_0}\alpha_s^2\left(0.07428-\ln
  g_s+1.9026g_s\right)$ \cite{CaronHuot:2007gq}. In our model, 
$m_T=0.8T$, then $g_s\approx1.6$, $\alpha_s\approx0.2$, and 
$\eta_D\approx \frac{0.9T^2}{M_0}$. To summarize, we make the Ansatz
\begin{eqnarray}
\eta_D =\frac{\gamma T^2}{M_0}\;,
\end{eqnarray} 
where $\gamma$ decreases with increasing 
$T$ for $T_c<T<3T_c$. Using a linear interpolation, 
we simply write
\begin{eqnarray}
\gamma=\frac{1}{a+bT/T_c} \;,
\end{eqnarray}
where $a$ and $b$ can be determined by the two limiting cases.

The obtained vector spectral function is shown in Fig.\
\ref{fig:rhoall}, where the shaded regions around the curves 
correspond to the uncertainties by halving or doubling the height of the
transport peak in the default model. Note that the uncertainties 
are relatively small. With increasing $T$, the resonance peak becomes 
broader and decreases in height, which indicates the dissociation of 
$J/\psi$. We estimate the dissociation temperature to be around
$2T_c$. On the other hand, the transport peak in the low-energy region
becomes higher with increasing $T$. Using Eq.\ \eqref{eq:hqd} and 
inserting the result for the QNS, we extract the $T$-dependence 
of the heavy quark diffusion coefficient. Furthermore, 
we can estimate $\eta/s$ according to Eq.\ \eqref{eq:etas}. 
The results are shown in Fig.\ \ref{fig:trans}. It is found that, for 
$1.5\,T_c<T<3.0\,T_c$, $\eta/s$ increases with $T$
and lies within the bounds given by the AdS/CFT 
and NLO-perturbative calculations, respectively.

In conclusion, combining a newly proposed spectral representation 
of the vector current correlation function with the self-consistent 
solutions of the quark gap equation and the inhomogeneous
Bethe-Salpeter equation, 
we calculated the charm quark diffusion coefficient $D$ of the QGP. 
Our result is consistent with that obtained by lattice QCD 
\cite{Ding:2012sp,Banerjee:2011ra} and that extracted by the PHENIX
experiment \cite{Adare:2010de}. With Eq.\ \eqref{eq:etas}, we
used $D$ to estimate $\eta/s$. We found that $\eta/s$ increases 
with increasing $T$. The values for $\eta/s$ remain above the 
AdS/CFT bound \cite{Kovtun:2003wp} and are close to that obtained from a
functional renormalization group calculation \cite{Haas:2013hpa} and other 
estimates (see Ref.~\cite{Adare:2010de} and references therein). 
In the future, we plan to extend the study to nonzero chemical
potential.

\section*{Acknowledgments}
S.-x.\ Qin would like to thank C.D.\ Roberts and Y.-x.\ Liu 
for helpful discussions. The work of S.-x.\ Qin was supported by the 
Alexander von Humboldt Foundation through a Postdoctoral Research Fellowship.

\end{document}